**Easy axis orientation of Strontium ferrite thin films described by spin reorientation**

P. Samarasekara and Udara Saparamadu

Department of Physics, University of Peradeniya, Peradeniya, Sri Lanka

## Abstract

In plane orientation of magnetic easy axis of sputtered strontium hexaferrite thin films has been explained using classical Heisenberg Hamiltonian. The variation of average value of in plane spin component with temperature was plotted in order to determine the temperature at which easy axis is oriented in the plane of the strontium ferrite film. The average value of in plane spin component in this 2-D model reaches zero at one particular temperature. This particular temperature obtained using our theoretical model agrees with the experimental value of the temperature of rf sputtered polycrystalline strontium ferrite thin films deposited on polycrystalline $Al_2O_3$ substrates (500 $^0$C). This spin reorientation temperature solely depends on the values of energy parameters used in our modified Heisenberg Hamiltonian equation.

## 1. Introduction:

Ferrite thin films are prime candidates in the application of magnetic memory devices and monolithic microwave integrated thin films (MMIC). Both strontium and barium ferrites belong to M-type hexagonal category. Due to its hard magnetic and uniaxial properties, hexagonal ferrites are unique among other ferrite materials. Strontium ferrite thin films have been synthesized on $Al_2O_3$ polycrystalline substrates using rf sputtering [1, 2], magnetron sputtering [3] and pulsed laser deposition [4]. The orientation of magnetic easy axis of ferrites vastly depends on the deposition or annealing temperature, orientation of the substrate and gas pressure inside deposition chamber.

Spin was assumed to be in the plane of y-z, and only two spin components ($S_y$ and $S_z$) were taken into account in this 2-D model. The average value of in plane ($S_y$) component was determined as a function of the temperature. By plotting $S_y$ versus temperature, the temperature at which $S_y$ approaches was investigated. Below this particular temperature ($T_s$), the magnetic easy axis orients in the plane of the film.

Different values of spin exchange interaction, long range magnetic dipole interaction, second order magnetic anisotropy, fourth order anisotropy and stress induced anisotropy were plugged in the equation of our modified Heisenberg Hamiltonian, in order to determine the variation of $T_s$. The out of plane easy axis orientation of barium ferrite thin films belonging to hexagonal ferrite has been previously explained by us using this model [5]. In addition, the easy axis orientations of soft spinel ferrite [6] and ferromagnetic [7] thin films have been explained previously. In all these cases, $S_y$ component was plotted against the temperature in order to investigate the orientation of magnetic easy axis. Furthermore, the total magnetic energy of Nickel ferrite and ferromagnetic films has been explained using unperturbed [11], the second order [8, 12, 16] and third order perturbed Heisenberg Hamiltonian [9, 14, 15]. According to our previous studies, stress induced anisotropy effects on coercivity [13]. However, the unperturbed Heisenberg Hamiltonian has been employed in this report.

## 2. Model:

The total energy of a magnetic thin film is given by following modified Heisenberg Hamiltonian [8, 9].

$$H = -J\sum_{m,n}\vec{S}_m.\vec{S}_n + \omega\sum_{m\neq n}\left(\frac{\vec{S}_m.\vec{S}_n}{r_{mn}^3} - \frac{3(\vec{S}_m.\vec{r}_{mn})(\vec{r}_{mn}.\vec{S}_n)}{r_{mn}^5}\right) - \sum_m D_{\lambda_m}^{(2)}(S_m^z)^2 - \sum_m D_{\lambda_m}^{(4)}(S_m^z)^4$$

$$-\sum_m \vec{H}..\vec{S}_m - \sum_m K_s Sin^2\theta_m \qquad (1)$$

Here *J, ω, θ,* $D_m^{(2)}, D_m^{(4)}, H_{in}, H_{out}, K_s$, *m* and *n* represent spin exchange interaction, strength of long range dipole interaction, azimuthal angle of spin, second and fourth order anisotropy constants, in plane and out of plane internal magnetic fields, stress induced anisotropy constant and spin plane indices, respectively. When the stress applies normal to the film plane, the angle between $m^{th}$ spin and the stress is $\theta_m$.

The long range magnetic dipole interaction of hexagonal ferrite calculated in one of our previous research articles has been used to find the total energy per unit spin given in following equation [5].

$E(\theta) = 3NJ+5(N-1)J+\omega[N(88.3197 \sin^2\theta+11.3541 \sin\theta\cos\theta -127.9435 \cos^2 \theta)$

$+(N-1)\ (93.0605 \sin^2\theta + 25.3002 \sin\theta \cos\theta - 15.423 \cos^2\theta)]$

$$-\cos^2\theta \sum_{m=1}^{N} D_m^{(2)} - \cos^4\theta \sum_{m=1}^{N} D_m^{(4)} + 3N(H_{in}\sin\theta + H_{out}\cos\theta + K_s \sin^2\theta) \qquad (2)$$

Here $\sum_{m=1}^{N} D_m^{(2)}$ and $\sum_{m=1}^{N} D_m^{(4)}$ represent the total second and fourth order anisotropy constants in the whole film.

Because only the $Fe^{+3}$ ions contribute to the net magnetic moment of hexagonal ferrites, the equation derived for barium ferrite can be applied for strontium ferrite too.

### 3. Results and Discussion:

The average value of in plane spin component is given by

$$\overline{S}_y = \frac{\int_0^{\pi} e^{-\frac{E}{kT}} \sin\theta d\theta}{\int_0^{\pi} e^{-\frac{E}{kT}} d\theta} \qquad (3)$$

Here *E*, *k* and *T* indicate the total magnetic energy given in equation (2), Boltzmann's constant and absolute temperature. Thickness of the strontium ferrite films incorporated for these simulations were approximately 2.5 μm thick. So value of N employed for these investigations was 998.

Figure 1 indicates the variation of $\overline{S}_y$ with temperature. When J = $10^{-33}$ Joules, $\omega=10^{-30}$ Joules, $\sum_{m=1}^{N} D_m^{(2)} = 10^{-29}$ Joules, $\sum_{m=1}^{N} D_m^{(4)} = 10^{-42}$ Joules, $K_s = 10^{-30}$ Joules, $H_{in}=10^{-32}$ $Am^{-1}$ and $H_{out}=10^{-39}$ $Am^{-1}$, $\overline{S}_y$ reaches zero at 773 K. This implies that in plane orientation of easy axis vanishes above 500 °C. Therefore, our experimental results of polycrystalline strontium ferrite thin film can be explained using this theoretical model [1]. The spin reorientation temperature ($T_s$) vastly depends on the energy parameters. When J = $10^{-33}$ Joules, $\omega=10^{-29}$ Joules, $\sum_{m=1}^{N} D_m^{(2)} = 10^{-29}$ Joules, $\sum_{m=1}^{N} D_m^{(4)} = 10^{-42}$ Joules, $K_s = 10^{-30}$ Joules, $H_{in}=10^{-32}$ $Am^{-1}$ and $H_{out}=10^{-39}$ $Am^{-1}$, $\bar{S}_y$ approaches zero at 160 K as shown in figure 2. This means that $T_s$ can be reduced by increasing ω.

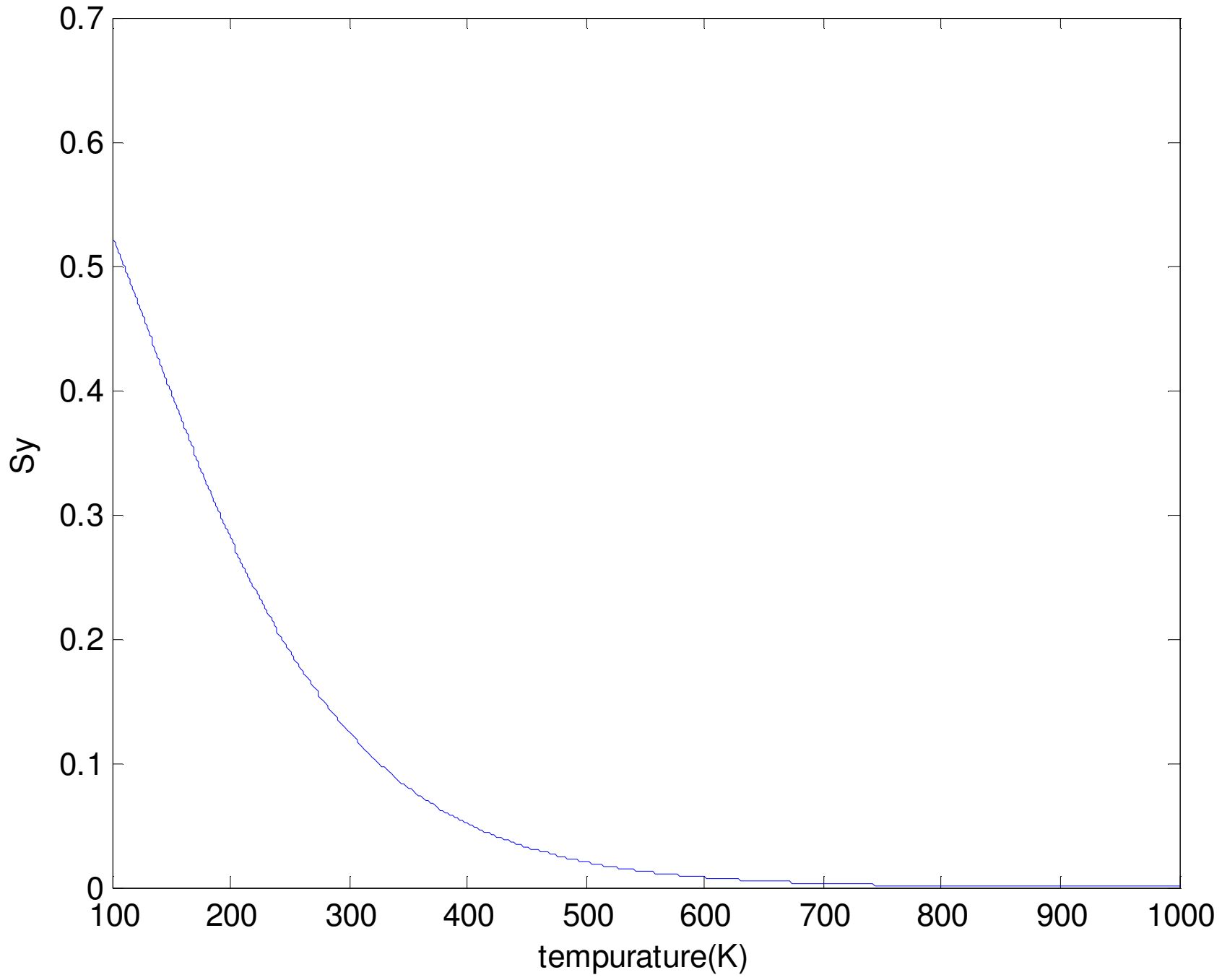

Fig 1: $\bar{S}_y$ versus temperature for the first set of values of energy parameters.

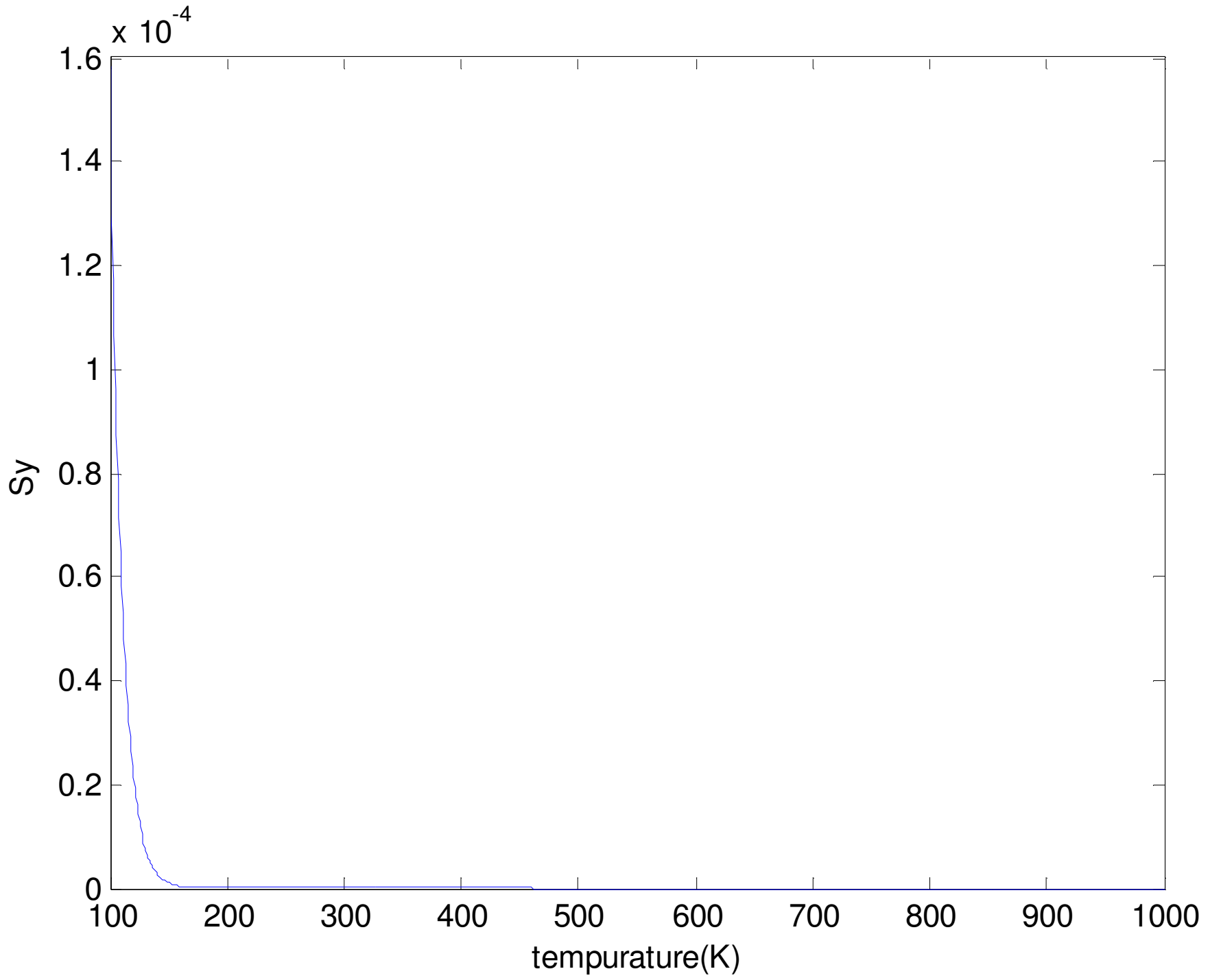

Fig 2: $\bar{S}_y$ versus temperature for the second set of values of energy parameters.

The variation of $T_s$ with J is given in figure 3. Other values of energy parameters were kept at $\omega$=$10^{-30}$ Joules, $\sum_{m=1}^{N} D_m{}^{(2)}$ =$10^{-29}$ Joules, $\sum_{m=1}^{N} D_m{}^{(4)}$ =$10^{-42}$ Joules, $K_s$= $10^{-30}$ Joules, $H_{in}$=$10^{-32}$ $Am^{-1}$ and $H_{out}$=$10^{-39}$ $Am^{-1}$ for this simulation. So $T_s$ gradually increases with J. Above J=$10^{-30}$ Joules, a rapid variation of $T_s$ can be observed. Below J=$10^{-31}$ Joules, $T_s$ doesn't vary with J. Between J=$10^{-30}$ and $10^{-31}$ Joules, $T_s$ slightly varies with J. Spin exchange interaction is related to the coupling between spins. So spins are restricted to rotate freely in a particular direction at higher values of J. As a result, $T_s$ increases with J.

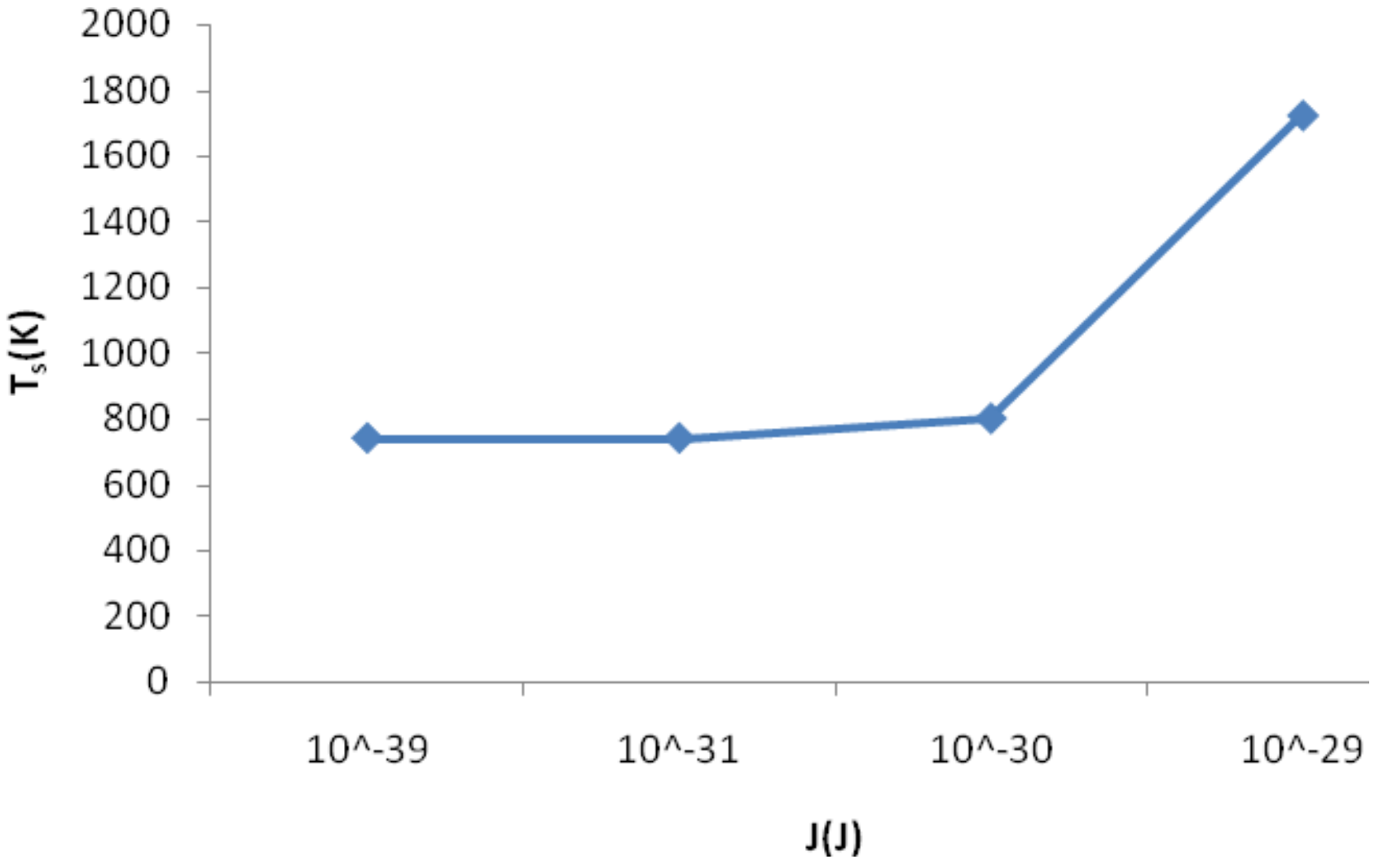

Fig 3: Variation of $T_s$ with J.

Figure 4 shows the variation of $T_s$ with $K_s$. Below $K_s=10^{-29}$ Joules, $T_s$ slightly varies with $K_s$. Above $K_s=10^{-29}$ Joules, $T_s$ drastically decreases with $K_s$. Other energy parameters were set to J = $10^{-33}$ Joules, $\omega=10^{-30}$ Joules, $\sum_{m=1}^{N} D_m^{(2)}=10^{-29}$ Joules, $\sum_{m=1}^{N} D_m^{(4)}=10^{-42}$ Joules, $H_{in}=10^{-32}$ $Am^{-1}$ and $H_{out}=10^{-39}$ $Am^{-1}$ in this simulation. Due to in plane stress, spins prefer to align in the in plane direction [10]. As a matter of fact, $T_s$ decreases with $K_s$. As shown in figure 5, $T_s$ varies with $H_{in}$. Other parameters were kept at J = $10^{-33}$ Joules, $\omega=10^{-30}$ Joules, $\sum_{m=1}^{N} D_m^{(2)}=10^{-29}$ Joules, $\sum_{m=1}^{N} D_m^{(4)}=10^{-42}$ Joules, $K_s$= $10^{-30}$ Joules and $H_{out}=10^{-39}$ $Am^{-1}$ for this simulation. Below $H_{in}=10^{-29}$ $Am^{-1}$, $T_s$ doesn't vary with $H_{in}$. Above $H_{in}=10^{-29}$ $Am^{-1}$, $T_s$ rapidly decreases with $H_{in}$. At larger values of internal in plane magnetic field, spins can easily rotate in the in plane direction. Then $T_s$ decreases with $H_{in}$.

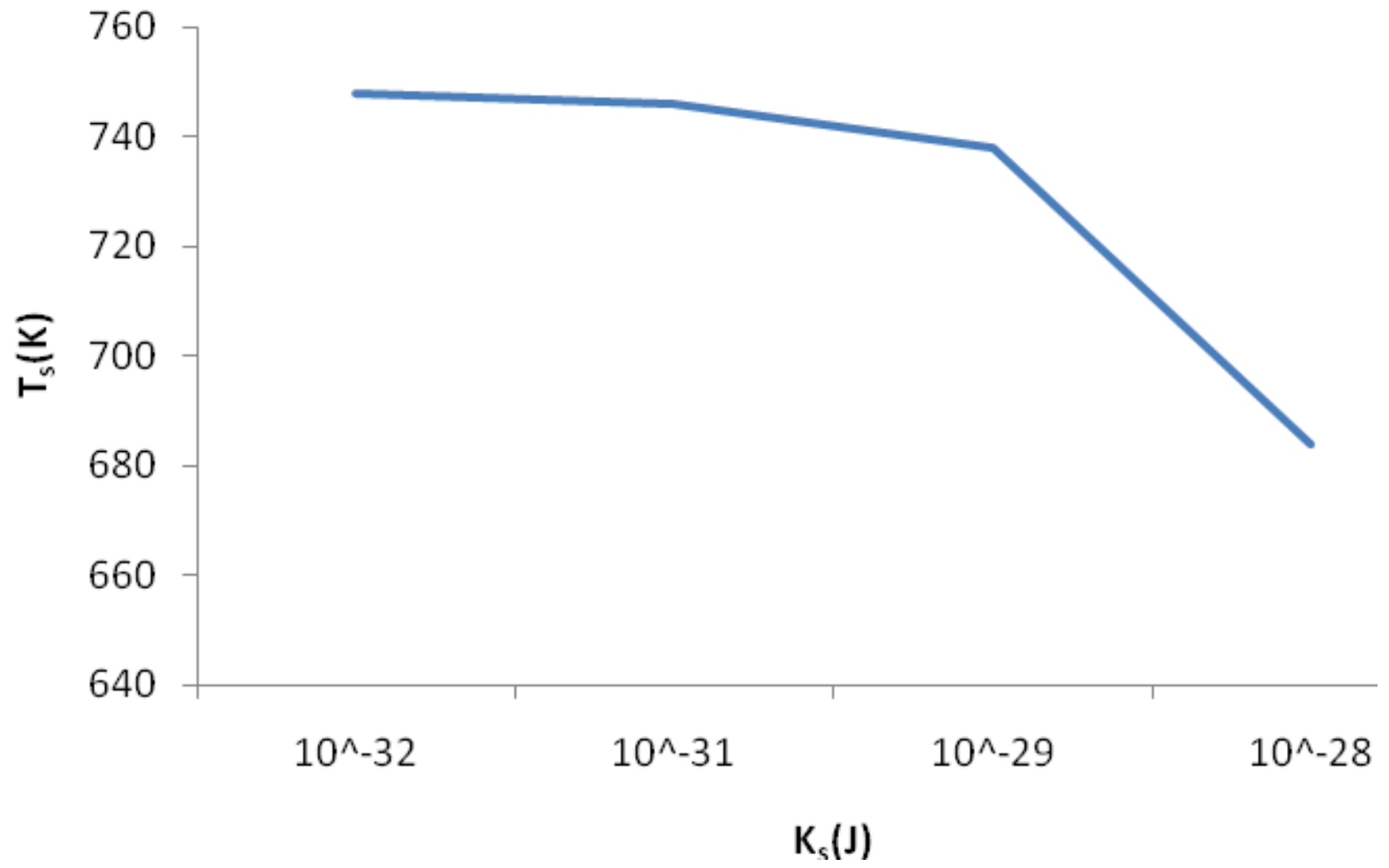

Fig 4: Plot of $T_s$ versus $K_s$.

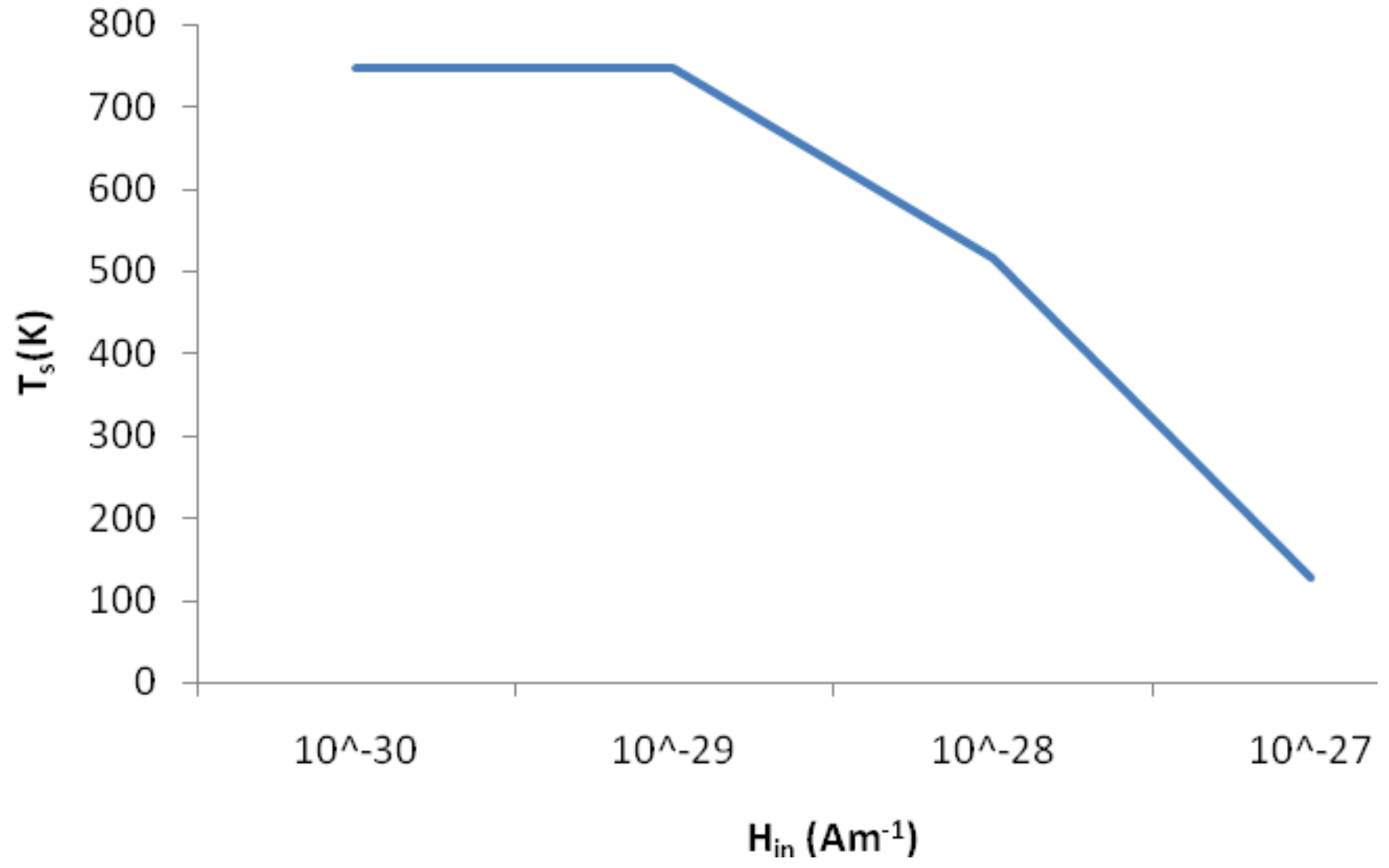

Fig 5: Dependence of $T_s$ on $H_{in}$.

**4. Conclusion:**

The easy axis orientation of polycrystalline strontium ferrite thin films sputtered on polycrystalline $Al_2O_3$ substrates could be explained using our modified Heisenberg Hamiltonian model. The total energy of oriented hexaferrite thin films derived from this model was employed in this case, rather than considering 2$^{nd}$ or 3$^{rd}$ order perturbation.

Variation of average value of in plane spin component with temperature was investigated. The spin reorientation temperature solely depends on ω, J, $K_s$ and $H_{in}$. However, $T_s$ is slightly sensitive to other energy parameters too. Below 500 °C, the easy axis of strontium ferrite is oriented in the plane of the film. This particular spin reorientation temperature could be obtained at J = $10^{-33}$ Joules, ω=$10^{-30}$ Joules, $\sum_{m=1}^{N} D_m^{(2)}$ =$10^{-29}$ Joules, $\sum_{m=1}^{N} D_m^{(4)}$ =$10^{-42}$ Joules, $K_s$= $10^{-30}$ Joules, $H_{in}$=$10^{-32}$ $Am^{-1}$ and $H_{out}$=$10^{-39}$ $Am^{-1}$. However, the spin reorientation temperature could be varied in a wide range by changing the values of J, ω, $\sum_{m=1}^{N} D_m^{(2)}$ , $\sum_{m=1}^{N} D_m^{(4)}$ , $K_s$, $H_{in}$ and $H_{out}$. Because the exact experimental values of J, ω, $\sum_{m=1}^{N} D_m^{(2)}$ , $\sum_{m=1}^{N} D_m^{(4)}$ , $K_s$, $H_{in}$ and $H_{out}$ of strontium thin films can't be found, a reasonable set of values has been employed for these explanations.

**References**

1. H. Hegde, P. Samarasekara and F.J. Cadieu, 1994. Nonepitaxial sputter synthesis of aligned strontium hexaferrites, SrO.6($Fe_2O_3$), films. J. Appl. Phys. 75(10), 6640-6642.
2. Antony Ajan, B. Ramamurthy Acharya, Shiva Prasad, S. N. Shringi and N. Venkataramani, 1998. Conversion electron Mössbauer studies on strontium ferrite films with in-plane and perpendicular anisotropies. J. Appl. Phys. 83, 6879.
3. A. Kaewrawang, A.Nagano Ghasemi, Xiaoxi Liu and A. Morisako, 2009. Properties of Sr Ferrite Thin Films on Al-Si Underlayer. IEEE trans. on Mag. 45(6), 2587-2589.
4. M.E. Koleva, S. Zotova, P.A. Atanasov, R.I. Tomov, C. Ristoscub, V. Nelea, C. Chiritescu, E. Gyorgy, C. Ghica and I.N. Mihailescu, 2000. Sr-ferrite thin films grown on sapphire by pulsed laser deposition. Applied Surface Science 168, 108-113.
5. P. Samarasekara and Udara Saparamadu, 2013. Easy axis orientation of Barium hexa-ferrite films as explained by spin reorientation. Georgian electronic scientific journals: Physics 1(9), 10-15.

6. P. Samarasekara and Udara Saparamadu, 2012. Investigation of Spin Reorientation in Nickel Ferrite Films. Georgian electronic scientific journals: Physics 1(7), 15-20.
7. P. Samarasekara and N.H.P.M. Gunawardhane, 2011. Explanation of easy axis Orientation of ferromagnetic films using Heisenberg Hamiltonian. Georgian electronic scientific journals: Physics 2(6), 62-69.
8. P. Samarasekara, 2010. Determination of Energy of thick spinel ferrite films using Heisenberg Hamiltonian with second order perturbation. Georgian electronic scientific journals: Physics 1(3), 46-52.
9. P. Samarasekara, 2011. Investigation of Third Order Perturbed Heisenberg Hamiltonian of Thick Spinel Ferrite Films. Inventi Rapid: Algorithm Journal 2, 1-3.
10. S. Chikazumi, Physics of magnetism, 1964, John Wiley & Sons., pages182-184.
11. P. Samarasekara, 2007. Classical Heisenberg Hamiltonian solution of oriented spinel ferrimagnetic thin films. Electronic Journal of Theoretical Physics 4(15), 187-200.
12. P. Samarasekara and S.N.P. De Silva, 2007. Heisenberg Hamiltonian solution of thick ferromagnetic films with second order perturbation. Chinese Journal of Physics 45(2-I), 142-150.
13. P. Samarasekara, 2003. A pulsed rf sputtering method for obtaining higher deposition rates. Chinese Journal of Physics 41(1), 70-74.
14. P. Samarasekara and William A. Mendoza, 2011. Third order perturbed Heisenberg Hamiltonian of spinel ferrite ultra-thin films. Georgian electronic scientific journals: Physics 1(5), 15-18.
15. P. Samarasekara and William A. Mendoza, 2010. Effect of third order perturbation on Heisenberg Hamiltonian for non-oriented ultra-thin ferromagnetic films. Electronic Journal of Theoretical Physics 7(24), 197-210.
16. P. Samarasekara, M.K. Abeyratne and S. Dehipawalage, 2009. Heisenberg Hamiltonian with Second Order Perturbation for Spinel Ferrite Thin Films. Electronic Journal of Theoretical Physics 6(20), 345-356.